\documentclass[11pt]{article}
\usepackage{graphicx}

\newcommand{\BABARPubYear}    {01}

\newcommand{\BABARProcNumber} {63}
\newcommand{\SLACPubNumber} {9040}

\newcommand{\babar}{$\mbox{\sl B\hspace{-0.4em} {\scriptsize\sl A}\hspace{-0.4em} B\hspace{-0.4em} {\scriptsize\sl A\hspace{-0.1em}R}}$}
\newcommand{\Lbabar}{\mbox{{\LARGE\sl B}\hspace{-0.15em}{\Large\sl A}\hspace{-0.07em}{\LARGE\sl B}\hspace{-0.15em}{\Large\sl A\hspace{-0.02em}R}}}
\newcommand{\lbabar}{\mbox{{\large\sl B}\hspace{-0.4em} {\normalsize\sl A}\hspace{-0.03em}{\large\sl B}\hspace{-0.4em} {\normalsize\sl A\hspace{-0.02em}R}}}
\newcommand{\pepii}{PEP-II}
\newcommand{\Bmeson}{$B$ meson}
\newcommand{\Bmesons}{$B$ mesons}
\newcommand{\Bdecays}   {$B$ decays}
\newcommand{\BF}{$B$ Factory}
\newcommand{\stwob}{\ensuremath{\sin\! 2 \beta}}
\newcommand{\result}{\ensuremath{\stwob=0.34\pm 0.20\, {\rm (stat)} \pm 0.05\, {\rm (syst)}}}
\newcommand{\CP}{\ensuremath{C\!P}}
\newcommand{\CPV}{\ensuremath{C\!P} violation}

\newcommand {\vub}{\ensuremath{|{V}_{ub}|}}
\newcommand {\vcb}{\ensuremath{|{V}_{cb}|}}

\newcommand {\vtd}{\ensuremath{|{V}_{td}|}}
\newcommand {\deltamd}{\ensuremath{{\rm \Delta}m_d}}
\newcommand {\mes}{\mbox{$m_{\rm ES}$}}
\newcommand {\mistag}{\ensuremath{w}}
\newcommand {\deltat}{\ensuremath{{\rm \Delta}t}}
\newcommand{\etaCP}{\ensuremath{\eta_{\CP}}}

\newcommand{\invfb}{\ensuremath{\mbox{\,fb}^{-1}}}
\newcommand{\invpb}{\ensuremath{\mbox{\,pb}^{-1}}}
\newcommand{\mev}{\ensuremath{\rm \,Me\kern -0.08em V}} 
\newcommand{\gev}{\ensuremath{\rm \,Ge\kern -0.08em V}} 
\newcommand{\gevc}{\ensuremath{{\rm \,Ge\kern -0.08em V\!/}c}} 
\newcommand{\mum} {\ensuremath{\,\mu\rm m}}

\newcommand{\dedx}      {\ensuremath{dE/dx}}
\newcommand{\ps}   {\ensuremath{\rm \,ps}}
\newcommand{\gevcc}{\ensuremath{{\rm \,Ge\kern -0.08em V\!/}c^2}}

\newcommand{\FourS} {\ensuremath{\Upsilon (4S)}}
\newcommand{\psitwos} {\ensuremath{\psi{(2S)}}}
\newcommand{\epem}{\ensuremath{e^+e^-}}
\newcommand{\Kb}    {\ensuremath{{\kern 0.18em\overline{\kern -0.18em K}}}{}} 
\newcommand{\Kz}    {\ensuremath{K^0}}

\newcommand{\B}{\ensuremath{B}}
\newcommand{\Bb}    {\ensuremath{{\kern 0.18em\overline{\kern -0.18em B}}}{}} 
\newcommand{\Bz}    {\ensuremath{B^0}}
\newcommand{\Bzb}   {\ensuremath{\Bb^0}}
\newcommand{\Bp}    {\ensuremath{B^+}}
\newcommand{\Bm}    {\ensuremath{B^-}}

\newcommand{\BB}    {\ensuremath{B\Bb}} 
\newcommand{\BzBzb} {\ensuremath{B^0 \Bb^0}}   
\newcommand{\jpsi}  {\ensuremath{{J\mskip -3mu/\mskip -2mu\psi\mskip 2mu}}}
\newcommand{\Ks}    {\ensuremath{K^0_{\scriptscriptstyle S}}}
\newcommand{\Kl}    {\ensuremath{K^0_{\scriptscriptstyle L}}}
\newcommand{\Kp}    {\ensuremath{K^+}}

\newcommand{\nim}       {Nucl.\ Instr.\ and Methods}
\newcommand{\jprlBase}  {Phys.\ Rev.\ Lett.}
\newcommand{\etal}{\hbox{\it et al.}{}}

\begin{document}
{\pagestyle{empty}

\begin{flushright}
SLAC-PUB-\SLACPubNumber \\
\babar-PROC-\BABARPubYear/\BABARProcNumber \\
October, 2001 \\
\end{flushright}

\par\vskip 4cm

\begin{center}
\Large \bf \CP\ violation in \Bdecays\ at the \Lbabar\ experiment
\end{center}
\bigskip

\begin{center}
\large 
P.J. Clark\\
Department of Physics, University of Bristol,
 Bristol, BS8 1TL, UK\\
(for the \lbabar\ Collaboration)
\end{center}
\bigskip \bigskip

\begin{center}
\large \bf Abstract
\end{center}
\noindent During the first year of data taking, the 1999-2000 run, the
\babar\ detector at the SLAC \pepii\ asymmetric collider has
collected an integrated luminosity of 20.7\invfb\ corresponding to 22.7
million \B\Bb\ pairs at the \FourS\ resonance. Using this data, we
present the measurement of \stwob\ based on samples of $\Bz \to \jpsi
\Ks$ , $\Bz \to \psitwos \Ks$ and $\Bz \to \jpsi \Kl$ decays. Our
measured value is \result. In addition we report on the measurement of
branching fractions for exclusive \Bdecays\ to charmonium final states,
measurements of charged and neutral \B\ meson lifetimes and also the
$\Bz\Bzb$ oscillation frequency.
\vfill
\begin{center}
Contributed to the Proceedings of the 24$^{th}$ International 
Workshop on Fundamental Problems in High Energy Physics and Field Theory, \\
6/27/2001---6/29/2001, Protvino, Moscow Region, Russia
\end{center}

\vspace{1.0cm}
\begin{center}
{\em Stanford Linear Accelerator Center, Stanford University, 
Stanford, CA 94309} \\ \vspace{0.1cm}\hrule\vspace{0.1cm}
Work supported in part by Department of Energy contract DE-AC03-76SF00515.
\end{center}

\section{Introduction}

The \babar\ experiment~\cite{tdr} is designed to allow detailed 
studies of \CPV\ in the \Bmeson\ system. The main goal is to investigate
whether \CPV\ can be fully explained within the Standard Model by the
imaginary phase of the Cabibbo-Kobayashi-Maskawa matrix, or if other
sources of \CPV\ should be considered~\cite{physbook}. To achieve this
goal we need to over constrain the Unitarity Triangle by measuring the
sides of the triangle using non-\CP\ violating processes which give
\vub, \vcb, and \vtd, while measuring the angles using \CP\ violating
processes~\cite{physbook}. The determination of the angles is achieved
by studying time dependent asymmetries of the neutral \Bmeson\ decays to
\CP\ eigenstates, $f_{CP}$,
\begin{eqnarray}
a_{f_{CP}}(t) & = & \frac{\Gamma(\Bz(t)\to f_{CP}) -
\Gamma (\Bzb(t)\to f_{CP})}{\Gamma(\Bz(t)\to f_{CP}) +
\Gamma (\Bzb(t)\to f_{CP})} \nonumber \\
& = & \frac{(1 - |\lambda|^2)\cos(\deltamd t) - 2\mathrm{Im}{\lambda}
\sin(\deltamd t) }{1+|\lambda|^2}.
\label{cpasym}
\end{eqnarray}
For certain \Bz\ decays, where other types of \CPV\ are assumed
negligible,  this equation reduces to
\begin{equation}
a_{f_{\CP}}(t)=-\mathrm{Im}(\lambda)\sin(\deltamd t),
\end{equation}
where $\lambda\equiv\frac{q}{p}\frac{\bar{A}_{f_{CP}}}{A_{f_{CP}}}$.
The complex parameters $q$ and $p$ are from \B\ mixing, \deltamd\ is a
measure of the $\Bz\Bzb$ oscillation frequency, and $\bar{A}_{f_{CP}}$
and $A_{f_{CP}}$ are the decay amplitudes (for more information
see~\cite{physbook}).  For some decay modes Im($\lambda$) is directly
related to angles of the Unitarity Triangle. In particular for the decay
$\Bz \to \jpsi \Ks$ we can obtain the angle $\beta$, given by
\begin{equation}
\mathrm{Im}(\lambda)=\stwob.
\end{equation}

\section{The \pepii\ storage ring}

The \pepii\ \BF~\cite{pepii:cdr} is an asymmetric-energy $e^+e^-$
collider which has a design luminosity of
3.0\,x10$^{33}$cm$^{-2}$s$^{-1}$ and a centre-of-mass energy, $\sqrt{s}$
= 10.58 GeV. This corresponds to the \FourS\ resonance which lies just
above the production threshold for $B_d$ mesons and below that for $B^*$
and $B_s$ mesons. The cross-sections for \FourS\ production and
continuum $q\bar{q}$, are 1.05\,nb and 3.39\,nb respectively.  The
branching ratio of the \FourS\ to \BB\ meson pairs is close to 100\%
providing around 3x10$^7$ \Bmeson\ pairs each year at design
luminosity. The small mass difference between the
$\Upsilon(\mathrm{4S})$ resonance and the two B mesons means that in the
$\Upsilon(\mathrm{4S})$ rest frame the B mesons are almost at rest and
travel only a short distance before decaying making their lifetimes hard
to measure. To overcome this an asymmetric machine was built. This uses
colliding beams of unequal energy to Lorentz boost the
$\Upsilon(\mathrm{4S})$ in the direction of the beam axis. The beam
energies chosen are 9\gev\ for the electrons and 3.1\gev\ for the
positrons ($\beta\gamma=0.56$). This results in the B mesons having
appreciable velocities and measurable decay lengths ($\sim$260\mum). At
the time of this conference the design luminosity had been achieved and
surpassed ($3.3\times10^{33}$). The design daily luminosity of
135\invpb\ per day was regularly achieved and exceeded with a maximum of
174.7\invpb\ achieved in one day.

\section{\babar\ detector}

A complete description of the \babar\ detector and its performance to
date can be found in Ref.~\cite{nim}. It consists of:
\begin{itemize}
\item a five layer double-sided silicon strip vertex tracker,
\item a cylindrical drift chamber filled with Helium-Isobutane
(80\%-20\%),
\item a \v{C}erenkov ring imaging particle identification system using
144 quartz bars,
\item a Caesium Iodide (CsI) electromagnetic calorimeter with 6580
crystals,
\item a superconducting solenoidal magnet (1.5 T),
\item an instrumented flux return with 19 layers of resistive plate chambers for muon
 identification and \Kl\ reconstruction. 
\end{itemize}

The tracking resolution can be parameterised
by $\sigma_{p_t} / p_t = (0.13\pm0.01)\%.p_t +
(0.45\pm0.03)\%$~\cite{nim}, where $p_t$ is in \gevc. The silicon vertex
tracker provides good vertex resolution with the $z$-resolution of the
\CP\ vertex typically being around 70\mum. Particle identification is
achieved using a combination of measurements from all \babar\
sub-detectors including the energy loss \dedx\ in the drift chamber and
in the silicon vertex tracker. Electrons and photons are identified in
the calorimeter, and muons are identified in the instrumented flux
return. The ring imaging \v{C}erenkov detector provides excellent
$\pi$-$K$ discrimination above 750\mev.

\section{Branching fraction measurements for \Bdecays\ to charmonium
final states} 

\begin{figure}[h!]
\includegraphics[width=75mm]{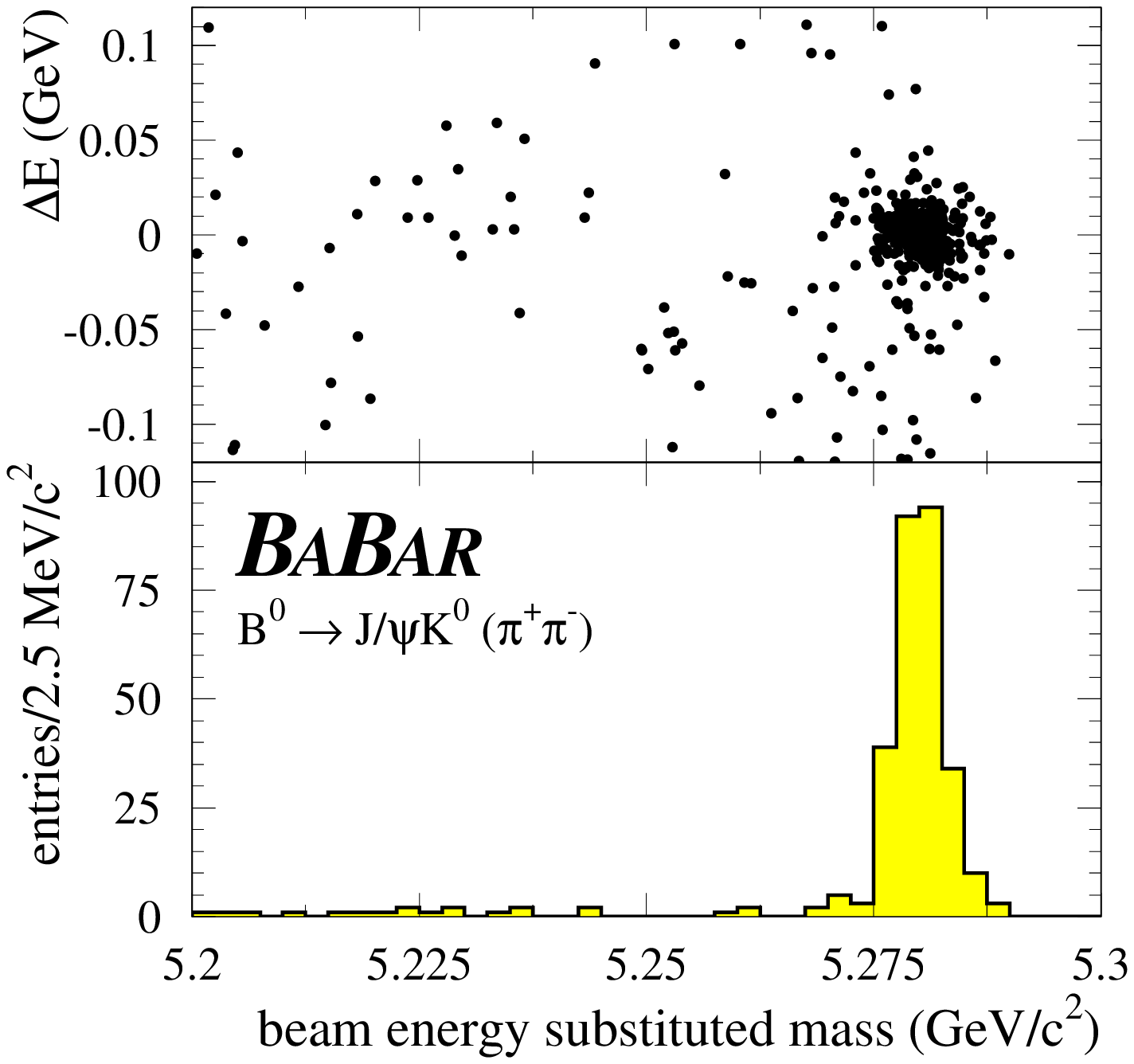}
\includegraphics[width=75mm]{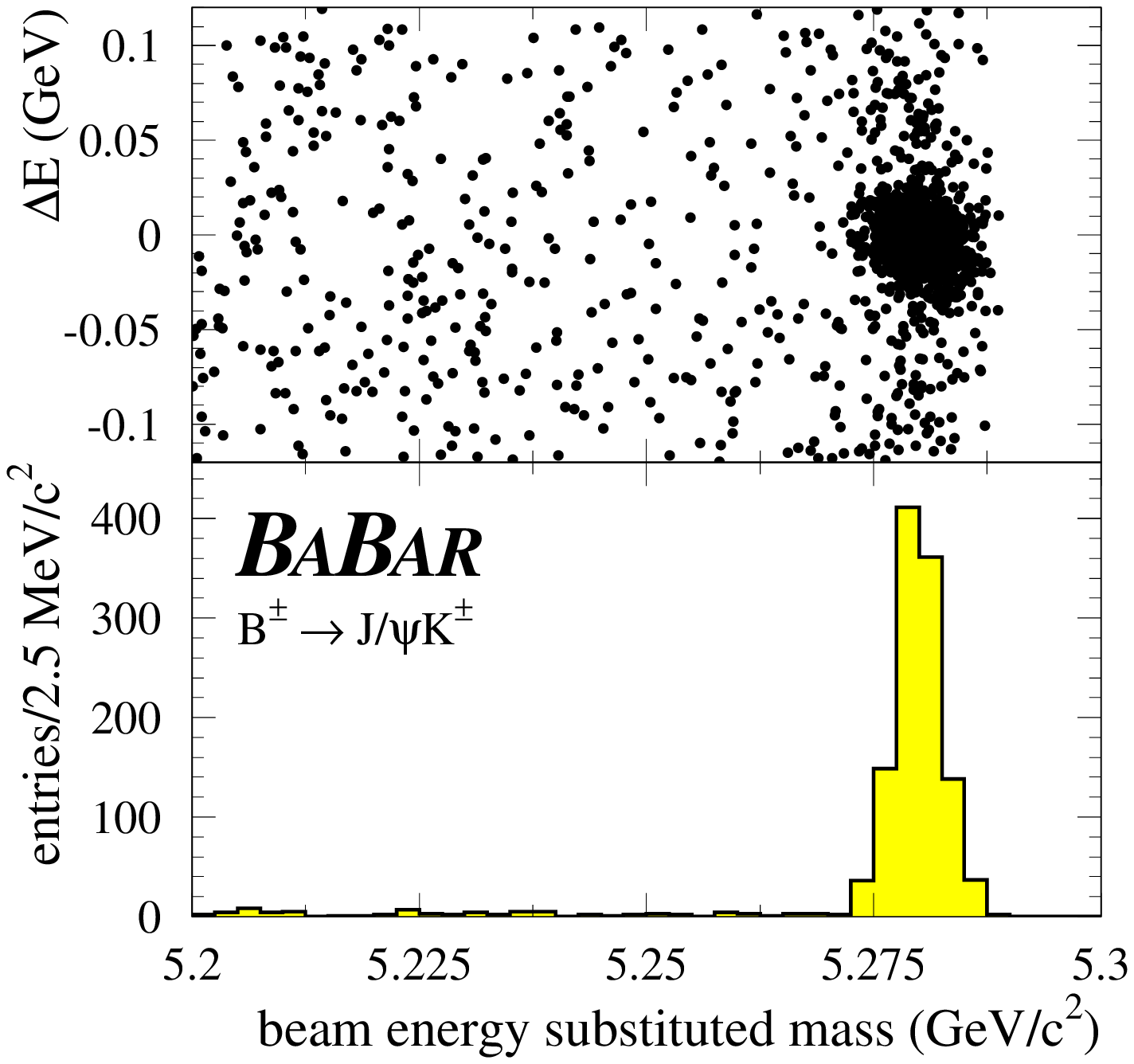}
\caption{\it Example distribution of the candidates in the $\Delta
E$--$m_{ES}$ plane for $\Bz\to\jpsi\Ks$ and $\Bp\to\jpsi\Kp$.}
\label{boxplots}
\end{figure}

Exclusive \Bmesons\ are identified using a variety of kinematic cuts. To
isolate the signal from background the variables $\Delta E$, the
difference between the reconstructed and expected \Bmeson\ energy in the
center-of mass frame, and $m_{ES}$, the beam energy substituted
mass are used. These variables are defined as follows:
\begin{eqnarray}
\Delta E & = & E_B^* - \frac{1}{2}\sqrt{s}, \\
 m_{ES}  & = & \sqrt{\left(\frac{1}{2}\sqrt{s}\right)^2-{P^*_B}^2}.
\end{eqnarray}

The resolution on $\Delta E$ is around 25\mev\ and is dominated by
detector resolution. The $m_{ES}$ resolution is around 3\mev\ and is
dominated by the beam energy spread.  Some example $\Delta E$ and
$m_{ES}$ distributions for two charmonium decays modes are shown in
Fig.~\ref{boxplots}. 

More detailed information on the measurement by \babar\ of the branching
fractions for \Bdecays\ to charmonium final states can be found
in~\cite{exclusive}. In the determination of the branching fractions we
have used the secondary branching fractions and their associated errors
published by the Particle Data Group~\cite{pdg}. The branching fractions
we have measured are shown in Table~\ref{charmonium}.
\begin{table}[h]
\begin{center}
\caption{\it Branching ratios to $c\bar{c}$ final states.}\vspace{2mm}
\begin{tabular}{|l|r|}\hline
Decay 				& \multicolumn{1}{l|}{Branching} \\ 
                                & \multicolumn{1}{l|}{Fraction (10$^{-4}$)} \\\hline
$\Bz \to \jpsi \Ks (\pi^+\pi^-)$  & $8.5 \pm 0.5 \pm 0.6$\\
$\Bz \to \jpsi \Ks (\pi^0\pi^0)$  & $9.6 \pm 1.5 \pm 0.7$\\
$\Bz \to \jpsi \Kl (\pi^+\pi^-)$  & $6.8 \pm 0.8 \pm 0.8$\\
$\Bz \to \jpsi \Kz (\mathrm{All})$         & $8.3 \pm 0.4 \pm 0.5$\\\hline
$\Bz \to \jpsi  K^{*0}$           & $12.4\pm 0.5 \pm 0.9$\\\hline
$\Bz \to \jpsi  K^{+}$            & $10.1\pm 0.3 \pm 0.5$\\\hline
$\Bp \to \jpsi  K^{*+}$           & $13.7\pm 0.9 \pm 1.1$\\\hline
$\Bz \to \chi_{c1} \Kz$           & $5.4 \pm 1.4 \pm 1.1$\\\hline
$\Bz \to \chi_{c1} K^{*0}$        & $4.8 \pm 1.4 \pm 0.9$\\\hline
$\Bz \to \chi_{c1} K^{+}$         & $7.5 \pm 0.8 \pm 0.8$\\\hline
$\Bz \to \psitwos \Kz$            & $6.9 \pm 1.1 \pm 1.1$\\\hline
$\Bz \to \psitwos  K^{+}$         & $6.4 \pm 0.5 \pm 0.8$\\\hline
$\Bz \to \jpsi \pi^{0}$           & $0.20\pm 0.06 \pm 0.02$\\\hline
\end{tabular}
\label{charmonium}
\end{center}
\end{table}

\section{\B\ lifetime measurement}

\begin{figure}[h]
\includegraphics[width=75mm]{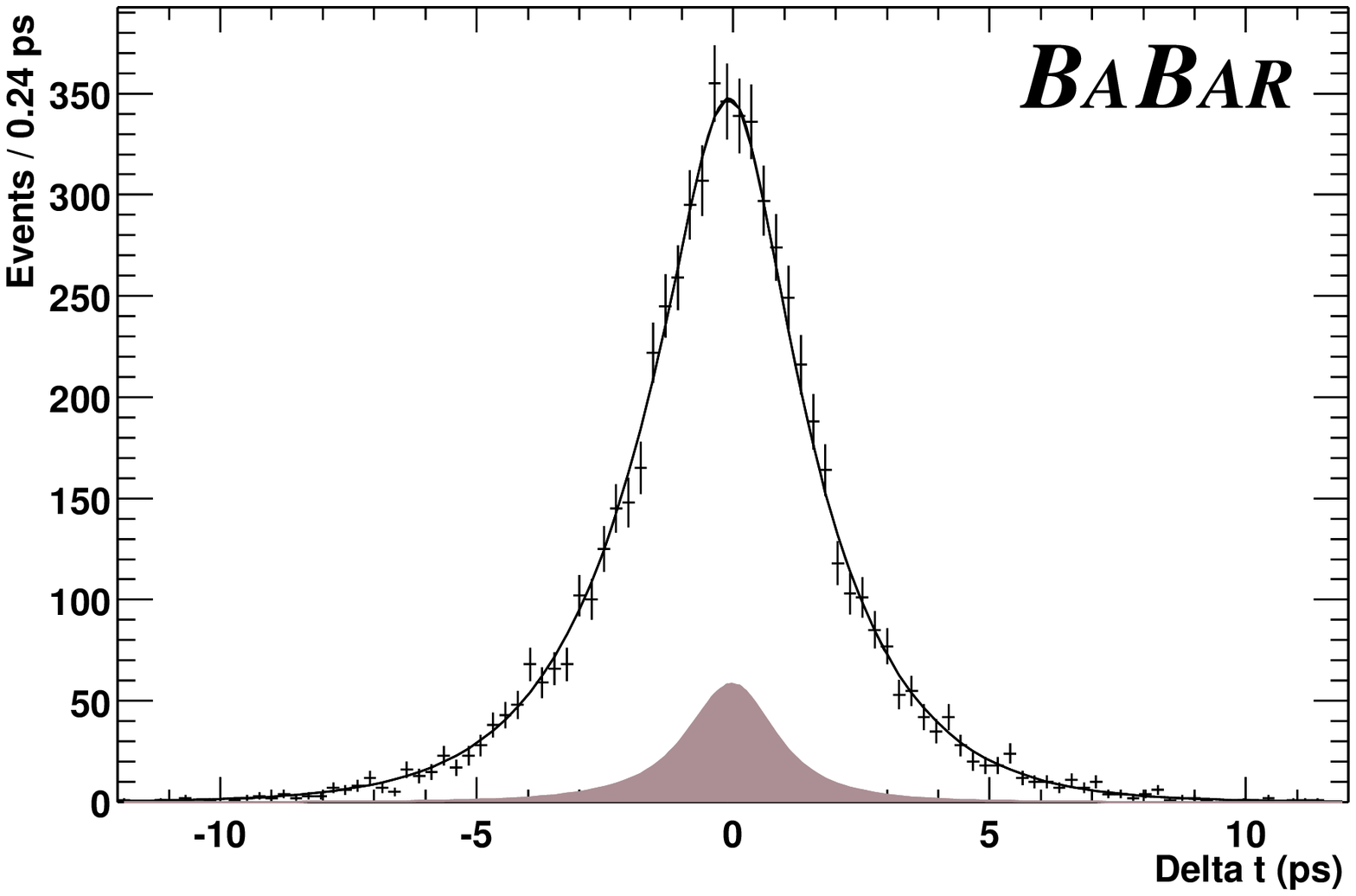}
\includegraphics[width=75mm]{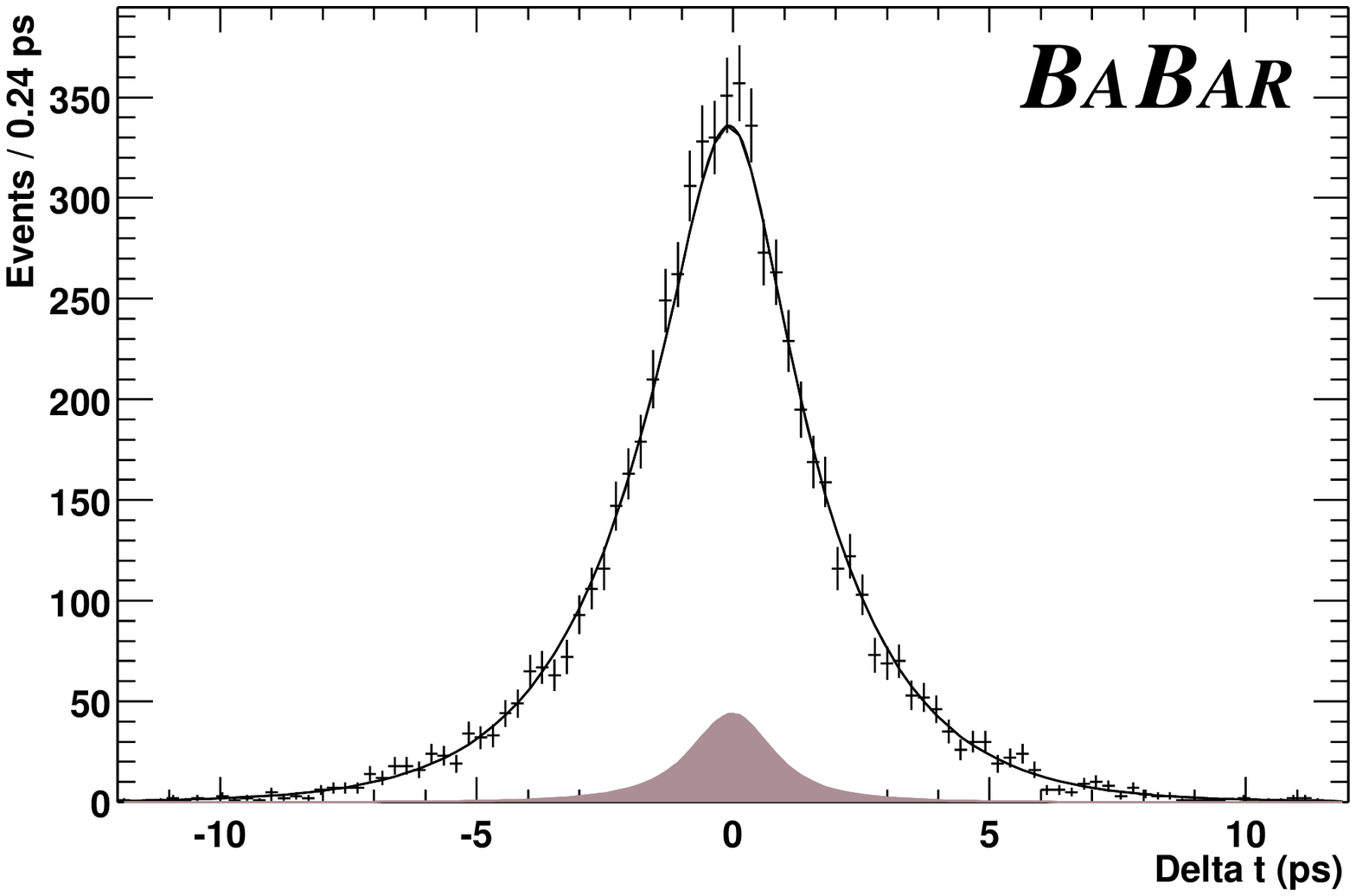}
\caption{\it $\Delta t$ distributions for \Bz/\Bzb\ (left) and \Bp/\Bm\
(right) candidates in the signal region (\mes$>5.27$ \gevcc). The result
of the lifetime fit is superimposed. The background is shown by the
hatched area.}
\label{fig:lifetime}
\end{figure}

Using the 20.7\invfb\ of data from run 1 it was possible to reconstruct
a very large number of charged and neutral B mesons. At the time of this
conference we have $6967\pm95$ neutral candidates and $7266\pm94$
charged candidates. This is one of the largest datasets of fully
reconstructed \Bmesons\ obtained to date. For complete information on
the measurement of the lifetimes see~\cite{PRLlifetimes}. The \Bz\ and
$B^{\pm}$ lifetimes are extracted from a simultaneous maximum unbinned
likelihood fit to the $\Delta t$ distributions of the signal candidates
where $\Delta t$ is obtained from measuring the distance $\Delta z$
between the decay vertices. To be able to perform this analysis it is
vital to understand and parameterize the detector vector resolution
function because the RMS resolution for measuring $\Delta t$ at \babar\
is around 1.3\ps\ which is significant when compared to the lifetimes
(1.5-1.6\ps). Studies with data and Monte Carlo simulation show that the
resolution function for \deltat\ is well modelled by the sum of a
zero-mean Gaussian distribution and its convolution with a decay
exponential shown in Eq.~\ref{resolfunct}, where $\delta_t$ is the
difference between the measured and true \deltat\ values, and $\hat
{a}$ represents the model parameters which are the fraction $h$ of
events in the core Gaussian component, a scale factor $s$ to take
account of the per-event errors $\sigma$, and a factor $\kappa$ in the
effective time constant of the exponential (for more information
see~\cite{PRLlifetimes}).
\begin{eqnarray}
{\cal {R}}\left(\delta_t,\sigma | \hat {a} = \left\{h,s,k\right\}\right)
 &=&  h\frac{1}{\sqrt{2\pi}s\sigma} \mathrm{exp}\left(-{\delta_t^2 \over 
2s^2\sigma^2}  \right)\nonumber \\  &+&  \int_{-\infty}^0 \frac{1-h}{\kappa\sigma}
\mathrm{exp}\left(\frac{\delta_t'}{\kappa\sigma}\right) \frac{1}{\sqrt{2\pi}s\sigma} \, {\rm exp} 
\left(  - {( \delta_t-\delta_{t}')^2 \over 
 2s^2\sigma^2 }  \right)\mathrm{d}(\delta_{t}').
\label{resolfunct}
\end{eqnarray}

Fig. \ref{fig:lifetime} shows the $\Delta t$ distributions with the
result of the lifetime fit superimposed. The preliminary results for the
\B\ meson lifetimes are:
\begin{eqnarray}
\tau_{\Bz} &=& 1.546 \pm 0.032\ {\rm (stat)} \pm 0.022\ {\rm
(syst)}\ps, \nonumber \\ \nonumber
\tau_{\Bp} &=& 1.673 \pm 0.032\ {\rm (stat)} \pm 0.022\ {\rm(syst)}\ps,\\
\tau_{\Bp}/\tau_{\Bz} &=& 1.082 \pm 0.026\ {\rm (stat)}\pm 0.011\ {\rm (syst).}\nonumber
\end{eqnarray}

\section{The \stwob\ measurement}

In \epem\ storage rings operating at the \FourS\ resonance the \BzBzb\
pairs produced in the \FourS\ decays evolve in coherent $P$-wave states
until one of the \B\ mesons decays.  If one of the \B\ mesons
($B_{tag}$) can be ascertained to decay to a state of known flavor at a
certain time $t_{tag}$, the other \B\ ($B_{CP}$) is {\it at that time}
known to be of the opposite flavor.

Each event which contains a $CP$ candidate is assigned a \Bz\ or \Bzb\
tag if the rest of the event satisfies some tagging criteria. The most
important parameter for tagging is the effective tagging quality
$Q=\epsilon(1-2\omega)^2$, where $\epsilon$ is the tagging
efficiency and $\omega$ is the probability of mis-tagging. The
statistical error on \stwob\ is inversely proportional to the square
root of the tagging quality. The efficiencies and probabilities of mistag
for the various types of tagging algorithm used are shown in
Table~\ref{tab:TagMix:mistag}. The algorithms are categorized in
four different types:
\begin{itemize}
\item lepton tag where the flavour of the \Bmeson\ is identified
using the charge of a high momentum lepton from a semileptonic decay,
\item kaon tag when the flavour of the \Bmeson\ is identified from
the charge of the kaon coming from the hadronisation of the $s$ quark
coming from the $b\to c \to s$ transitions,
\item NT1 and NT2 tags chosen using the output of a neural
network using correlated information such as secondary lepton
charge, slow pions from $D^*$ decays, and jet charge.
\end{itemize}
\begin{table}[h]
\caption{\it Mistag fractions measured from a maximum-likelihood fit to
the time distribution for the fully-reconstructed \Bz\ sample. The uncertainties on $\varepsilon$ and $Q$ are statistical only.} 
\begin{center}
\begin{tabular}{|l|r|l|r|}  \hline 
Category     & \multicolumn{1}{c}{ $\varepsilon$
(\%)}&\multicolumn{1}{c|}{$\mistag$ (\%)} & \multicolumn{1}{c|}{$Q$ (\%)}\\ \hline 
{\tt Lepton} & 10.9 $\pm$ 0.4 & $11.6\pm2.0$ &   $6.4\pm0.7$  \\ 
{\tt Kaon}   & 36.5 $\pm$ 0.7 & $17.1\pm1.3$ &  $15.8\pm1.3$  \\ 
{\tt NT1}    & 7.7  $\pm$ 0.4  & $21.2\pm2.9$ &   $2.6\pm0.5$  \\ 
{\tt NT2}    & 13.7 $\pm$ 0.5 & $31.7\pm2.6$ &   $1.8\pm0.5$  \\  \hline 
All          & 68.9 $\pm$ 1.0 &              &  $26.7\pm1.6$  \\  \hline 
\end{tabular} 
\end{center}
\label{tab:TagMix:mistag}
\end{table}

For the measurement of \stwob\ , $B_{CP}$ is fully
reconstructed in a \CP\  eigenstate ($\jpsi \Ks$, $\psitwos \Ks$ or
$\jpsi \Kl$). By measuring the proper time interval $\deltat = t_{CP} -
t_{tag}$ from the $B_{tag}$ decay time to the decay of the $B_{CP}$
($t_{CP}$), it is possible to determine the time evolution of the
initially pure \Bz\ or \Bzb\ state: 
\begin{equation}
\label{eq:TimeDep}
        f_\pm(\, \deltat \, ; \,  \Gamma, \, \deltamd, \, {\cal {D}} \sin{2\beta } )  = {\frac{1}{4}}\, \Gamma \, {\rm e}^{ - \Gamma \left| \deltat 
\right| }\, \left[  \, 1 \, \mp \, {\cal {D}}\etaCP \sin{ 2 \beta } \times \sin{\deltamd \, \deltat } \,  \right],
\end{equation}
where the $+$ or $-$  sign  indicates whether the 
$B_{tag}$ is tagged as a \Bz\ or a \Bzb, respectively.  The dilution
factor ${\cal {D}}$ is given by $ {\cal {D} } = 1 - 2 \mistag$, where
$\mistag$ is the mistag fraction, {\it i.e.}, the  probability that the
flavor of the tagging \B\ is identified incorrectly.  \etaCP\ is the CP
eigenstate of the final state and it is $\etaCP=-1$ for the $\jpsi \Ks$
and  $\psitwos \Ks$ modes, $\etaCP=+1$ for the $\jpsi \Kl$ mode. Although less pure, the $\jpsi \Kl$ mode is very important because the oscillation is expected to be opposite to the other ones. 

To account for the finite resolution of the detector,
the time-dependent distributions $f_\pm$ for \Bz\ and \Bzb\ tagged events 
(Eq.~\ref{eq:TimeDep}) must be convoluted with 
the time resolution function mentioned earlier ${\cal {R}}( \deltat ; \hat {a} )$:
\begin{equation}
\label{eq:Convol}
        {\cal F}_\pm(\, \deltat \, ; \, \Gamma, \, \deltamd, \, {\cal {D}}\etaCP
 \sin{ 2 \beta }, \hat {a} \, )  = 
f_\pm( \, \deltat \, ; \, \Gamma, \, \deltamd, \, {\cal {D}}\etaCP \sin{ 2 \beta
 } \, ) \otimes 
{\cal {R}}( \, \deltat \, ; \, \hat {a} \, ),
\end{equation}

Extraction of the amplitude of the \CP\ asymmetry and the value of
\stwob\ is done with an unbinned maximum likelihood fit.  In order to
extract as much information from the data itself and properly account
for correlation, the fit is performed simultaneously to the CP and the
flavor eigenstates. There are 35 parameters free in the fit. For the fit
the \Bz\ lifetime and \deltamd\ are fixed to the currently best known
values~\cite{pdg}. The \deltat\ distributions for \Bz\ and \Bzb\ tags
are shown in Fig.~\ref{deltatcp}. The results of the fit for \stwob\,
using the full tagged data sample are shown below. More information and
the latest results can be found in~\cite{PRL1,PRL2}.
\begin{equation}
\result.
\end{equation}
\begin{figure}[htb]
\begin{center}
  \includegraphics[width=60mm]{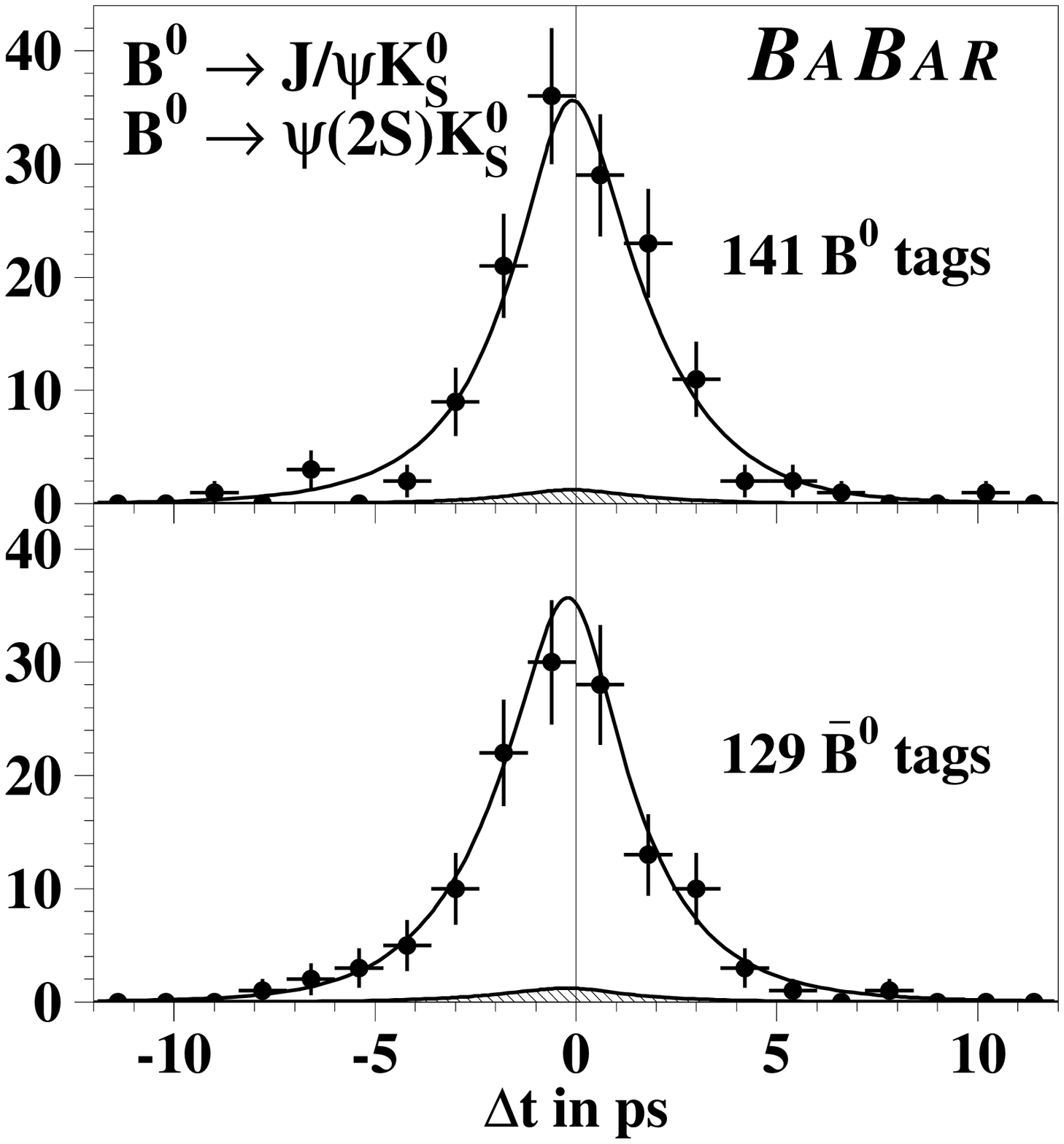}
  \includegraphics[width=60mm]{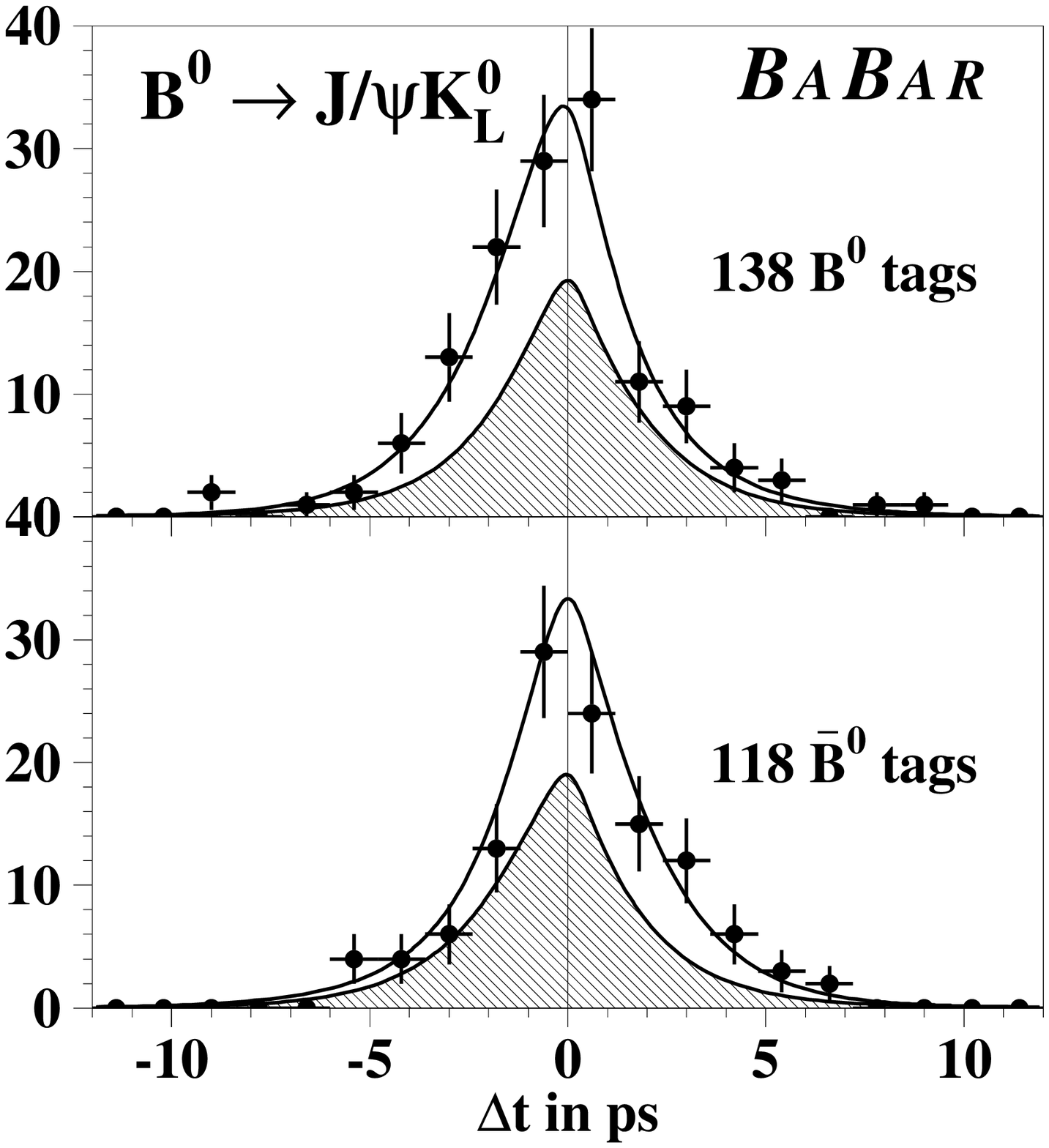}
\caption{\it Distribution of \deltat\ for the \Bz\ and \Bzb\ tags. Shown
on the left for  $\Bz \to \jpsi \Kz$ and $\Bz \to
\psitwos \Ks$, and on the right for $\Bz \to \jpsi \Kl$. The background
which is significantly larger for the \Kl\ mode is shown by the hatched area.}
\label{deltatcp}
\end{center}
\end{figure}

\section{$B$ mixing}

The time dependent \Bz\Bzb\ mixing measurement requires the
determination of the flavour of both \Bmesons. Considering the \Bz\Bzb\
system as a whole, one can classify the tagged events as \emph{mixed} or
\emph{ummixed} depending upon whether the $B$s are tagged with the same
flavour or opposite flavour. From the time-dependent rate of mixed
($N_{mix}$) and unmixed ($N_{unmix}$) events, the mixing asymmetry
$a(\Delta t) = (N_{unmix}-N_{mix})/(N_{unmix}+N_{mix})$ is calculated as
a function of $\Delta t$ and fitted to the expected cosine distribution.
Fig.~\ref{fig:mixing} shows the \deltat\ and $a(\Delta t)$ distributions
with the fit results superimposed.
\begin{figure}[htb]
\begin{center}
  \includegraphics[width=70mm]{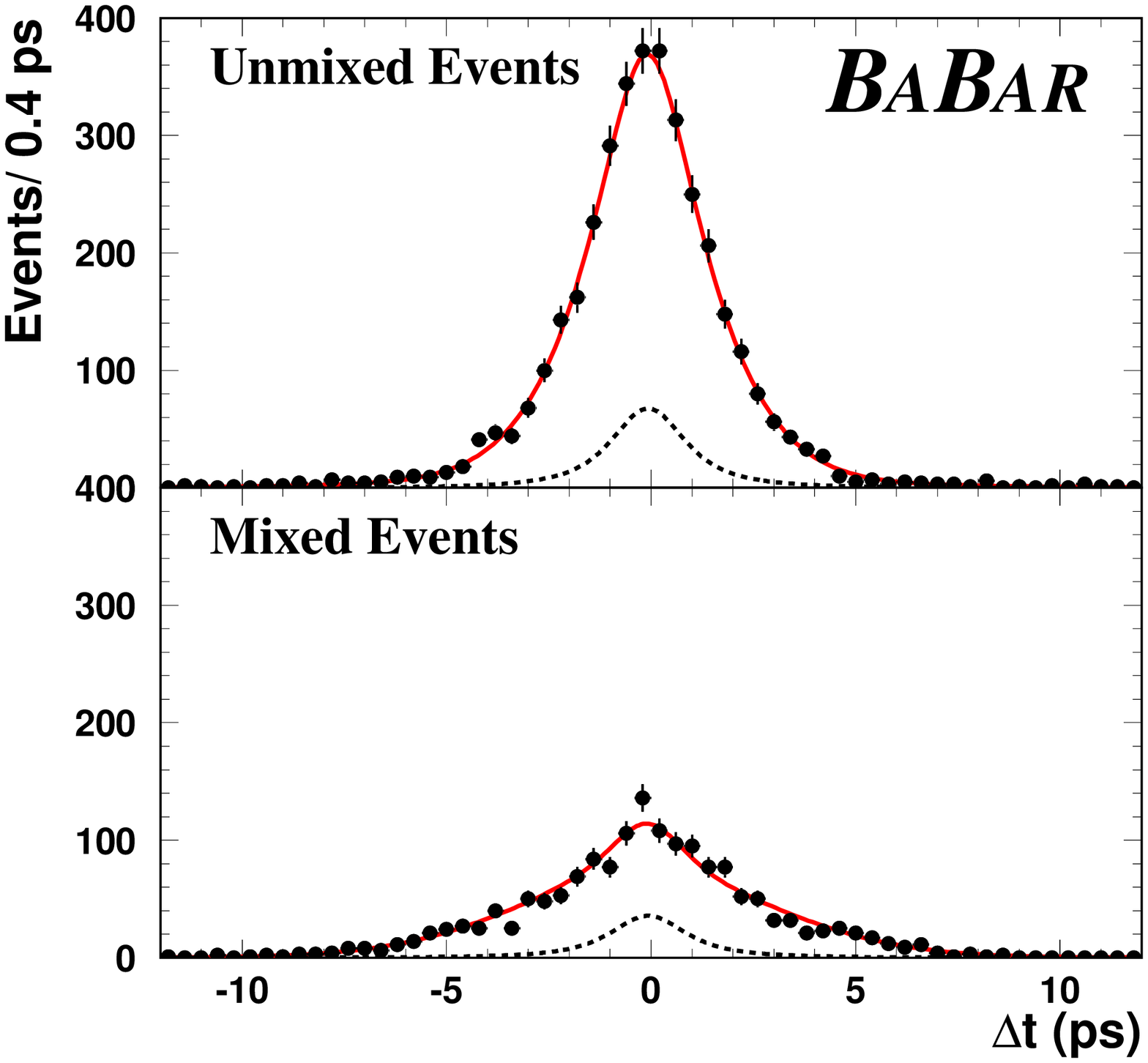}
  \includegraphics[width=85mm]{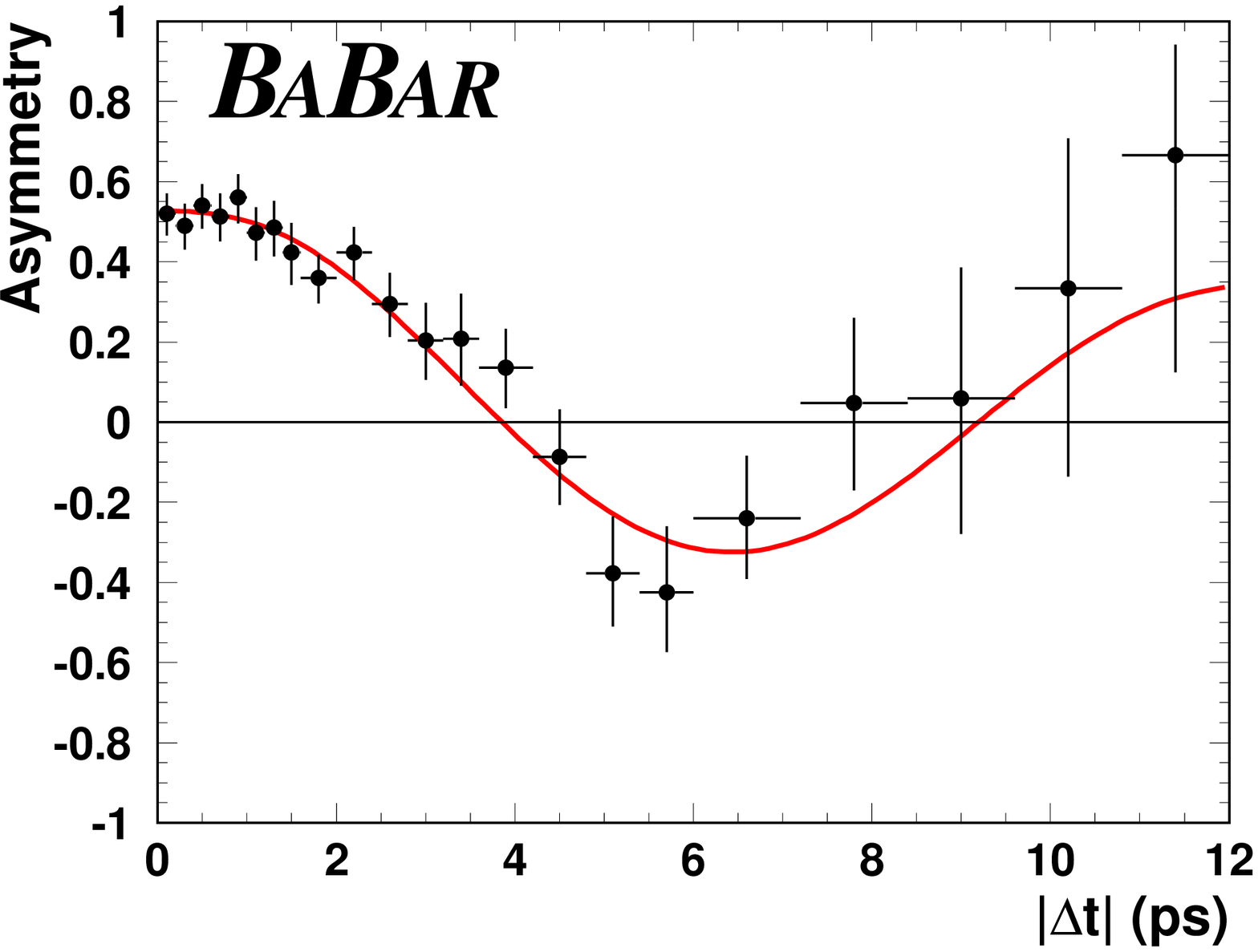}
\caption{\it \deltat\ distribution for mixed and unmixed events (left)
and the time-dependent asymmetry $a(\Delta t)$ between unmixed and mixed events (right).}
\label{fig:mixing}
\end{center}
\end{figure}

We measure the \BzBzb\ oscillation frequency to be
\begin{equation}
\deltamd  =  0.519 \pm 0.020\ ({\rm stat})  \pm 0.016  ({\rm
  syst})\  \hbar {\rm ps}^{-1}.
\end{equation}
This result is consistent with previous measurements~\cite{pdg}
and is of similar precision.

\section{Conclusion}

We have presented \babar 's  measurement of the \CP\ violating asymmetry
parameter \stwob\ in the \Bmeson\ system:
\begin{equation}
\result.
\end{equation}
The measurement is consistent with the world average $\stwob =
0.9\pm0.4$~\cite{pdg},  and it is currently statistically limited by the
size of the \CP\ sample. 

We have also presented measurements of various branching fractions for
\Bdecays\ to charmonium final states, $B$ lifetime measurements, and
time--dependent mixing which has been performed for the first time at
the \FourS.

\end{document}